\documentclass[aps, epsf, amsfonts]{iopart}
\usepackage{graphics}
\usepackage{graphicx}
\usepackage{epsfig}

\def\beq{\begin{equation}}
\def\eeq{\end{equation}}
\def\bea{\begin{eqnarray}}
\def\eea{\end{eqnarray}}
\def\hp{h_{+}}
\def\hc{h_{\times}}

\def\fp{F^{+}}
\def\fc{F^{\times}}

\begin{document}
\input epsf.tex

\title{Detecting galactic binaries with LISA.}
\author{Neil J. Cornish and Edward K. Porter}
\address{Department of Physics, Montana State University, Bozeman, MT 59719, USA.}
\ead{porter@physics.montana.edu}
\vspace{1cm}
\begin{abstract}
\noindent One of the main sources of gravitational waves for the LISA space-borne interferometer are galactic binary systems.  The waveforms for these sources are represented by eight parameters of which four are intrinsic and four are extrinsic to the system.  Geometrically, these signals exist in an 8-d parameter space.  By calculating the metric tensor on this space, we calculate the number of templates needed to search for such sources.  We show in this study that below a particular monochromatic frequency of $f_{0}\sim 1.6\times 10^{-3}$ Hz we can ignore one of the intrinsic parameters and search over a 7-d space.  Beyond this frequency, we have a change in dimensionality of the parameter space from 7 to 8 dimensions.  This sudden change in dimensionality results in a change in the scaling of template number as a function of the monochromatic frequency from $\sim f_{0}^{1.25}$ to $\sim f_{0}^{5.88}$.
\end{abstract}

\maketitle

\section{Introduction}
One of the main sources of gravitational waves for the planned space-borne Laser Interferometer Space Antenna (LISA) are quasi-monochromatic galactic binary systems.  The LISA detector is composed of three spacecraft which form an equatorial triangle and will work in the frequency range $10^{-5} \leq f/Hz\leq 1$.  The center of mass of the constellation traces out a circular orbit around the Sun at a distance of 1 AU and lies about $20^{o}$ behind the Earth.  The three spacecraft cartwheel in retrograde motion as they move around the Sun with a period of one year.  This motion induces amplitude, frequency and phase modulations in the gravitational wave signal\cite{cutler,cornishlarson}.  The amplitude modulation is caused by the antenna pattern being moved across the sky.  With the three arms, LISA acts like a pair of two-arm detectors.  The constellation is set up such that the arm lengths of the detector are equal with an angle of $\pi/3$ between each detector arm.  This is equivalent to the response of two $90^{o}$ interferometers rotated by an angle of $\pi/4$ radians with respect to each other. Because LISA can be thought of as two separate detectors, measuring different polarizations of the GW, the phase modulations are cause by combinations of the polarizations.  Finally, the frequency or Doppler modulations are caused by the motion of the detector with respect to the source.  The noise in the LISA detector is complicated by the inclusion of a transfer function at high frequencies.  Due to this, we will work, in this study, in the Low Frequency Approximation (LFA).  In the LFA we can set the transfer function due to the arm length of the detector equal to unity for $f \ll f_{*}=10^{-2}$ Hz.

In order to detect galactic binaries one can employ the method of matched filtering
~\cite{Helst}.  Briefly, the method works as follows: Firstly, one creates a set of waveforms, or templates as they are called, that depend on a number of parameters of the source and its location and orientation relative to the detector. These templates are then cross-correlated with the detector output weighted by the inverse of the noise spectral density.  If a signal, whose parameters are close to one of the template waveforms, is actually present in the
detector output then the cross-correlation builds up, with the dominant contribution coming from frequencies where the noise spectral density is low.  Thus, in the presence of a sufficiently strong signal the correlation will be much larger than the RMS correlation in the
absence of any signal. How large should it be before we can be confident about the presence of
a signal depends on the combination of the number of sources and the false alarm probability.  The effectiveness of matched filtering depends on how well the phase evolution of the waveform is known. Even tiny instantaneous differences, as low as one part in $10^3$ in the phase of the true signal that might be present in the detector output and the template that is used to dig it out could lead to a cumulative difference of several radians since one integrates over several hundreds to several thousands of cycles.

Using a geometric method~\cite{owen}, we can treat the parameters of the systems as a set of basis coordinates.  We can then define a scalar product and metric tensor in this vector space.  Having done this we can show that we only need to carry out our search over a lower dimensional sub-space.  The search for galactic binaries will then involve the placement of a grid of templates in this sub-space.

\section{The Gravitational Waveform.} \label{sec:waveform}
The strain of the gravitational wave with both polarizations is
\beq
h(t) = \hp\,\fp + \hc\,\fc,
\eeq
where the two polarizations are given by
\beq
h_{+}\left(t\right) = A_{0}\left(1+\cos^{2}\iota\right)\cos\left(\Phi\left(t\right)-\varphi_{0}\right), 
\eeq
\beq
h_{\times}\left(t\right) = -2\,A_{0}\cos\iota\,\sin\left(\Phi\left(t\right)-\varphi_{0}\right),
\eeq
where $A_{0}$ is a constant initial amplitude, $\iota$ the angle of inclination of the source, and $\varphi_{0}$ is a constant initial phase.  The phase of the gravitational wave, $\Phi(t)$, for a circular orbit is defined by
\beq
\Phi(t) = 2\pi f_{0}t + \pi \dot{f_{0}} t^{2} + 2\pi\left(f_{0}+\dot{f_{0}}t\right)R\,\sin(\theta)\cos\left(2\pi f_{m}t-\phi\right),
\eeq
where $f_{0}$ is the quasi-monochromatic frequency and $\dot{f_{0}}$ is its first time derivative given by
\begin{equation}
\frac{df_{0}}{dt}=\frac{96}{5}\pi^{8/3}{\mathcal M}_{c}^{5/3}f_{0}^{11/3}.
\end{equation}
Here, ${\mathcal M}_{c}=m\eta^{3/5}$, is the chirp mass, and, $m=m_{1}+m_{2}$ and $\eta=m_{1}m_{2}/m^{2}$ are the total mass and reduced mass ratio.  ($\theta,\phi$) define the location of the source, $f_{m} = 1/T$ is the modulation frequency and $T$ is the number of seconds in a year, and $R$ is the light travel time for 1 AU ($\sim$ 500 secs).  The quantities $F^{+}$ and $F^{\times}$ are defined in the LFA by~\cite{rubbo}
\beq
F^{+}(t;\psi, \theta, \phi, \lambda) = \frac{1}{2}\left[\cos(2\psi)D^{+}(t;\theta, \phi, \lambda) - \sin(2\psi)D^{\times}(t;\theta, \phi, \lambda)\right],
\eeq
\beq
F^{\times}(t;\psi, \theta, \phi, \lambda) = \frac{1}{2}\left[\sin(2\psi)D^{+}(t;\theta, \phi, \lambda) + \cos(2\psi)D^{\times}(t;\theta, \phi, \lambda)\right],
\eeq
where $\psi$ is the polarization angle of the wave and $\lambda = 0$ or $3\pi/2$ defines the two-arm combination of LISA from which the strain is coming.  The detector pattern functions are given by
\bea
D^{+}(t) = \frac{\sqrt{3}}{64}\left[\frac{}{}-36\sin^{2}(\theta)\sin(2\alpha(t)-2\lambda)+(3+\cos(2\theta)) \right.\\ \nonumber\fl \left(\frac{}{}\cos(2\phi)\left\{\frac{}{}9\sin(2\lambda)-\sin(4\alpha(t)-2\lambda)\right\} \frac{}{}+\sin(2\phi)\left\{\frac{}{}\cos\left(4\alpha(t)-2\lambda\right)-9\cos(2\lambda) \right\}\frac{}{}\right)\\ \nonumber  \left.-4\sqrt{3}\sin(2\theta)\left(\frac{}{}\sin(3\alpha(t)-2\lambda-\phi)-3\sin(\alpha(t)-2\lambda+\phi)\right)\right]
\eea
\bea
D^{\times}(t) = \frac{1}{16}\left[\frac{}{}\sqrt{3}\cos(\theta)\left(\frac{}{}9\cos(2\lambda-2\phi)-\cos(4\alpha(t)-2\lambda-2\phi) \right) \right. \\ \nonumber \left. -6\sin(\theta)\left(\frac{}{} \cos(3\alpha(t)-2\lambda-\phi)+3\cos(\alpha(t)-2\lambda+\phi) \right) \right],
\eea
where $\alpha(t)=2\pi t / T$ is the orbital phase of the center of mass of the constellation.

\section{The geometric method.}
The set of ${\mathcal N}$ parameters which define the system can also be thought of as defining an ${\mathcal N}$ dimensional manifold.  This template manifold shares the properties of a Hilbert space.  On this manifold is it possible to define a scalar product of the two time series waveforms $h(t)$ and $g(t)$ as
\begin{equation}
\left< h| g\right> = \int_{-\infty}^{\infty}\frac{df}{S_{n}(f)} \tilde{h}(f)\tilde{g}^{*}(f) + c.c.,
\label{eq:sp}
\end{equation}  
where the tilde denotes the Fourier transform $\tilde{h}(f)=\int_{-\infty}^\infty h(t) \exp(2\pi i f t) dt$ of the time series waveform $h(t)$, a * denotes complex conjugate and the {\emph c.c.} denotes a complex conjugate term.  The quantity $S_{n}(f)$ represents the two-sided noise spectral density of the detector, which is given by~\cite{cornish}
\begin{equation}
S_{n}(f)=4 S_{n}^{pos}(f)+8\left(1+\cos^{2}\left(\frac{f}{f_{*}}\right) \right)\frac{S_{n}^{accel}(f)}{(2\pi f)^{4}} , 
\end{equation}
where $S_{n}^{pos}(f)$ and $S_{n}^{accel}(f)$ are the position and acceleration noise respectively.  As we are searching for monochromatic binaries, we can greatly simplify the calculation of the inner product.  In this case, the noise at the quasi-monochromatic frequency can be treated as a constant and taken outside the integral.  Then, due to Parseval's theorem, we can calculate the scalar product in the time domain.  This simplifies Equation~(\ref{eq:sp}) to 
\begin{equation}
\left< h| g\right> = \frac{2}{S_{n}(f_{0})}\int_{0}^{T}dt\,h(t)g(t),
\label{eq:sp2}
\end{equation}  
where $T$ is the period of integration (which in this study we will assume to be $T$ = 1 year).
If we consider unnormalized waveforms, we can write the metric tensor on this manifold in the form
\begin{equation}
g_{\mu\nu}=\frac{\left<\partial_{\mu}h\left|\partial_{\nu}h\right>\right.}{\left<h\left|h\right>\right.}-\frac{\left<h\left|\partial_{\mu}h\right>\right.\left<h\left|\partial_{\nu}h\right>\right.}{\left<h\left|h\right>\right.^{2}}.
\end{equation}
The first term on the right hand side is recognized as being the Fisher information matrix.  The extra term on the right hand side arises from the fact that we have not pre-normalized the waveforms.  The number of templates needed to carry out a gravitational wave search is simply the proper volume of the parameter space, divided by the proper volume of the template.  Assuming a D-dimensional hyper-cubic template, the number of templates is given by
\begin{equation}
{\mathcal N}=\frac{\int d^{\,D}\lambda \sqrt{g}}{dl^{D}},
\end{equation}
where $g=det|g_{\mu\nu}|$ is the determinant of the metric tensor, $\lambda$ is the parameter set which defines the manifold and $dl^{D}$ is the proper volume of the template.  While the galactic binaries are defined by an 8-d parameter set, we only need to search over four intrinsic parameters.  The reason for this will be explained in the next section.  In order to do this, we project from the initial 8-d space to the 4-d search sub-space using the recursive projection
\begin{equation}
\gamma_{ij}=g_{ij}-\frac{g_{ki}g_{kj}}{g_{{kk}}},
\end{equation}
which carries out a projection onto the sub-space orthogonal to the $k$-dimension. It can be shown that this projection onto orthogonal sub-spaces is equivalent to maximizing over the projected parameter~\cite{ekp}.  Using the fact that the proper distance between two templates is related to a pre-determined threshold, called a minimum match, $MM$,via
\begin{equation}
MM=1-\frac{1}{2}ds^{2}=1-\frac{1}{2}g_{\mu\nu}d\lambda^{\mu}d\lambda^{\nu},
\end{equation}
we can now write the proper volume of the template in $D$ dimensional space as
\begin{equation}
dl^{D}=\left(2\sqrt{\frac{1-MM}{D}}\right)^{D}.
\label{eqn:tempdist}
\end{equation}

\section{Results.}
For numerical purposes it simplifies things if we use the parameters $p=f_{0}T$ and $q(p)=\dot{f_{0}}T^{2}$.  For quasi-monochromatic binaries, the initial parameter space is defined by the set $\lambda^{\mu}=\{A_{0}, \imath, \varphi_{0}, \psi, \theta, \phi, p, q\}$.  The first four of these parameters are kinematic or extrinsic parameters.  The four parameters $\{\theta, \phi, p, q\}$ are dynamical parameters dependent on the astrophysical system in question and are called intrinsic parameters (the parameters $(\theta,\phi)$ are considered intrinsic due to the motion of the detector with respect to the source). 
 
Instead of having to search over the full 8-d search space, we only need to search of the sub-set of intrinsic parameters $\bar{\lambda}^{\mu}=\{\theta,\phi,p,q\}$.  This is due to the fact that we can use the F-statistic to find the optimal values of the four extrinsic parameters.  This is done as follows.  We first write the strain of the gravitational wave in the form
\begin{equation}
h(t) = \sum_{i=1}^{4}a_{i}\left(A_{0}, \imath, \varphi_{0}, \psi\right)A^{i}\left(t;\theta,\phi,p,q\right).
\end{equation}
We then define four constants $N^{i}=\left<s \left| A^{i}\right>\right.$, where $s$ is the signal, and find a solution for the $a_{i}$'s in the form
\begin{equation}
a_{i} = \left(M^{-1}\right)_{ij}N^{j},
\end{equation}
where
\begin{equation}
M^{ij} = \left<A^{i} \left| A^{j}\right>\right. .
\end{equation}
If we now substitute the above into the expression for the reduced log-likelihood, i.e.
\begin{equation}
\ln \,{\mathcal L}\left(\lambda^{\mu}\right) = \left<s \left| h\left(\lambda^{\mu}\right)\right>\right. - \frac{1}{2}\left<h\left(\lambda^{\mu}\right) \left| h\left(\lambda^{\mu}\right)\right>\right. ,
\end{equation}
we obtain the F-statistic~\cite{JKS}
\begin{equation}
{\mathcal F} = \frac{1}{2}\left(M^{-1}\right)_{ij}N^{i}N^{j},
\end{equation}
which automatically maximizes the log-likelihood over the extrinsic parameters and reduces the search space to the sub-space of intrinsic parameters.
 
\begin{figure}[t]
\begin{center}
\epsfig{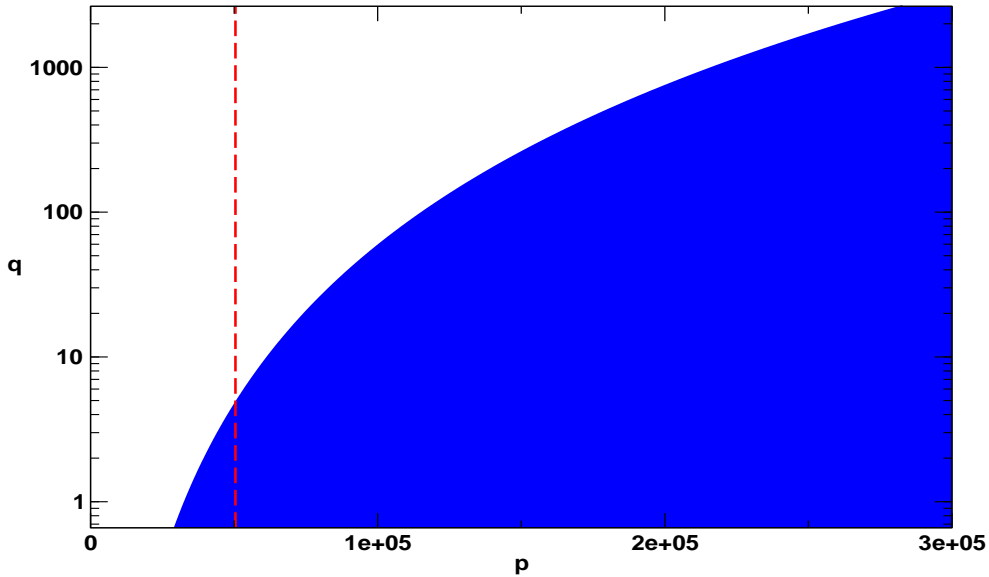}
\end{center}
\caption{A log-linear plot of $p$ against $q(p)$ for galactic binaries.  During the Monte Carlo integration we sample from the shaded area, where the boundary of this area corresponds to a maximum chirp mass of 20 $M_{\odot}$.  The dashed line corresponds to the value of $p$ beyond which we have to increase the dimensionality of the search space and consider the parameter $q$.}
\label{fig:pvsq}
\end{figure}

In calculating the number of templates needed, we take different limits of the parameter $p$, we set a maximum value of $q$ corresponding to a maximum chirp mass of ${\mathcal M}_{c}=20\,M_{\odot}$ (giving us a distribution of sample values of $q$ for each value of $p$) and use these as a bounding box for a Monte Carlo integration to calculate the proper volume of the search manifold.  As the boundary of our search space is quite regular, we can approximate the proper volume of the search space as
\begin{equation}
\int d^{\,{\mathcal N}}\lambda\,\sqrt{g}=\frac{V\sum_{i}\sqrt{g_{i}}}{N},
\end{equation}
where $V = \Delta\theta\Delta\phi\Delta p\Delta q$ is the volume of the search space, $\sqrt{g_{i}}$ is the value of the metric determinant at a particular point, and $N$ is the total number of points used.  We can see from Figure~(\ref{fig:pvsq}) that choosing a maximum value of $q(p)$ for each $p$ provides us with a range of values to sample over during the Monte Carlo integration.  At the same time as carrying out the Monte Carlo integration, we also compare the proper distance in the $q$-dimension for each value of $p$, i.e.
\begin{equation}
s_{q} = \int_{0}^{q_{max}(p)}dq\, \sqrt{g_{qq}}, 
\end{equation}
with the proper distance between two templates in the $q$-dimension, $dl$, as defined by Equation~\ref{eqn:tempdist}.  It is only when $s_{q}\geq dl$ that we need to take this parameter into account and increase the dimensionality of the initial search manifold from 7-d to 8-d.  Increasing the dimensionality of the search manifold infers a corresponding increase in the dimensionality of the sub-manifold of intrinsic parameters from 3-d to 4-d.  This increase happens at $p^{*}\sim 5\times 10^{4}$ and we now have $s_{q} \sim dl$.  This transition point is indicated by the dashed line in Figure~(\ref{fig:pvsq}).

\begin{figure}[tb]
\begin{center}
\epsfig{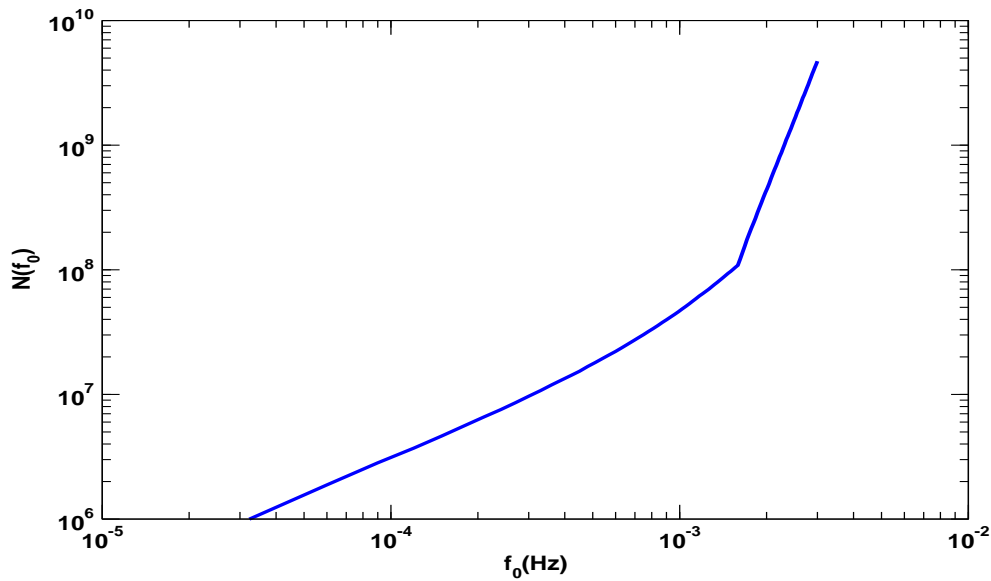}
\end{center}
\caption{The cumulative number of templates needed to search for individual quasi-monochromatic galactic binaries up to a particular value of $f_{0}$ with the LISA interferometer.  We assume in this figure that the observation period $T$ = 1 year.}
\label{fig:LISATemplates}
\end{figure}
In Figure~(\ref{fig:LISATemplates}) we have plotted the cumulative number of templates, ${\mathcal N}(f_{0})$, needed to carry out a matched filtering search for individual quasi-monochromatic galactic binaries up to a particular choice of $f_{0}$.  We have assumed a minimal match in this case of $MM=0.97$.  We can see from the plot that at a frequency of $f_{0}=f_{0}^{*}\sim 1.6\times 10^{-3}$ Hz, there is a point of inflection and a sudden increase in the growth of template number.  It is because of the increase in dimensionality that we have the sudden increase in the growth of template number.  We should also point out that the position of the change of dimensionality is dependent on our choice of $MM$.  A higher choice of $MM$ will push the point corresponding to the change in dimensionality to the left on our graph.

In terms of how the number of templates scales against frequency, this increase in dimensionality causes a big change.  Below $f^{*}_{0}$, the number of templates scales as $\sim f_{0}^{1.25}$.  After the change in dimensionality, this changes to $\sim f_{0}^{5.88}$.

\section{Conclusions}\label{sec:conclusions}
In this study we have calculated the number of templates needed to carry out the search for individual quasi-monochromatic galactic binaries with the LISA interferometer.  Using the standard geometric method, we have treated the waveform parameters as the basis coordinates in an 8-d space.  As most of the parameters are extrinsic to the system, we only need to carry out the search on a 4-d sub-manifold.  We find that below a frequency of $f^{*}_{0}=1.6\times 10^{-3}$ Hz, the proper distance between two templates in the $\dot{f_{0}}$ direction is greater than the proper coordinate distance in that direction.  As a consequence, below this frequency we can completely ignore the parameter defining this direction and our initial search space reduces to 7-d.  It is only above  $f^{*}_{0}$ that the proper coordinate distance in that direction becomes equal to or greater than the proper distance between two templates.  We then have an increase in the dimensionality of the search space from 7-d to 8-d.  It is because of this increase in dimensionality that we have the increase in the scaling law from $\sim f_{0}^{1.25}$ to $\sim f_{0}^{5.88}$.

\end{document}